\documentclass[preprint,journal]{vgtc}       

\ifpdf
  \pdfoutput=1\relax                   
  \usepackage{graphicx}                
  \DeclareGraphicsExtensions{.pdf,.png,.jpg,.jpeg} 
\else
  \usepackage{graphicx}                
  \DeclareGraphicsExtensions{.eps}     
\fi%

\graphicspath{{figures/}{pictures/}{images/}{./}} 

\usepackage{microtype}                 
\PassOptionsToPackage{warn}{textcomp}  
\usepackage{textcomp}                  
\usepackage{mathptmx}                  
\usepackage{times}                     
\usepackage{cite}                      
\usepackage{tabu}                      
\usepackage{booktabs}                  

\usepackage{xcolor}
\usepackage{paralist}
\usepackage{amssymb}
\usepackage{float}
\usepackage{placeins}

\usepackage{scrextend}

\DeclareMathAlphabet{\mathcal}{OMS}{cmsy}{m}{n}

\usepackage[draft=true]{minted}
\usepackage[hyphens]{url}
\usepackage[hidelinks]{hyperref}
\hypersetup{breaklinks=true}

\usepackage{multirow}

\newenvironment{rev}[0]{%
    \leavevmode\color{black}\ignorespaces
}{}

\newcommand{\mvf}{MobileVisFixer}

\definecolor{class}{RGB}{117, 161, 210}
\definecolor{gp}{RGB}{56, 115, 192}
\definecolor{encoding}{RGB}{158, 88, 209}
\definecolor{lDep}{RGB}{56, 151, 101}

\newcommand{\tClass}[1]{\textcolor{class}{\mintinline{html}{#1}}}
\newcommand{\tGroup}[1]{\textcolor{gp}{\mintinline{html}{#1}}}
\newcommand{\tEncoding}[1]{\textcolor{encoding}{\mintinline{html}{#1}}}
\newcommand{\tLdeps}[1]{\textcolor{lDep}{\mintinline{html}{#1}}}

\newcommand{\tLdep}[2]{\textcolor{lDep}{\mintinline{html}{#1(}}\textcolor{gp}{\mintinline{html}{#2}}\textcolor{lDep}{\mintinline{html}{)}}
}

\newcommand{\formatHTML}[1]{\mintinline{html}{#1}}

\usepackage{amsmath}

\usepackage{array}

\usepackage[ruled,vlined]{algorithm2e}
\SetKwComment{Comment}{$\triangleright$\ }{}

\newcommand{\bpstart}[1]{\vspace{1mm} \noindent{\textbf{#1.}}}

\onlineid{1067}

\vgtccategory{Research}
\vgtcpapertype{algorithm/technique}

\title{\mvf{}:~Tailoring Web Visualizations for Mobile Phones Leveraging an Explainable Reinforcement Learning Framework}

\author{Aoyu Wu, Wai Tong, Tim Dwyer, Bongshin Lee, Petra Isenberg, and Huamin Qu}
\authorfooter{
\item
 A. Wu, W. Tong, and H. Qu are with Hong Kong University of Science and Technology. Email: $\{$awuac, wtong, huamin$\}$@ust.hk.
\item
 T. Dwyer is with Monash University. E-mail: Tim.Dwyer@monash.edu.
\item
 B. Lee is with Microsoft Research. E-mail: bongshin@microsoft.com.
 \item 
 P. Isenberg is with Inria. Email: petra.isenberg@inria.fr.
}

\shortauthortitle{Biv \MakeLowercase{\textit{et al.}}: Global Illumination for Fun and Profit}

\abstract{We contribute \mvf{}, a new method to make visualizations more mobile-friendly. 
Although mobile devices have become the primary means of accessing information on the web, many existing visualizations are not optimized for small screens and can lead to a frustrating user experience. 
Currently, practitioners and researchers have to engage in a tedious and time-consuming process to ensure that their designs scale to screens of different sizes, and existing toolkits and libraries provide little support in diagnosing and repairing issues.
To address this challenge, \mvf{} automates a mobile-friendly visualization re-design process with a novel reinforcement learning framework.
To inform the design of \mvf{}, we first collected and analyzed SVG-based visualizations on the web, and identified five common mobile-friendly issues.
\begin{rev}
\mvf{} addresses four of these issues on single-view Cartesian visualizations with linear or discrete scales by a Markov Decision Process model that is both generalizable across various visualizations and fully explainable. 
\end{rev}
\mvf{} deconstructs charts into declarative formats, and uses a greedy heuristic based on Policy Gradient methods to find solutions to this difficult, multi-criteria optimization problem in reasonable time. 
\begin{rev}
In addition, \mvf{} can be easily extended with the incorporation of optimization algorithms for data visualizations. 
\end{rev}
Quantitative evaluation on two real-world datasets demonstrates the effectiveness and generalizability of our method.
} 


\CCScatlist{ 
 \CCScat{K.6.1}{Management of Computing and Information Systems}%
{Project and People Management}{Life Cycle};
 \CCScat{K.7.m}{The Computing Profession}{Miscellaneous}{Ethics}
}

\teaser{
  \centering
  \includegraphics[width=\linewidth]{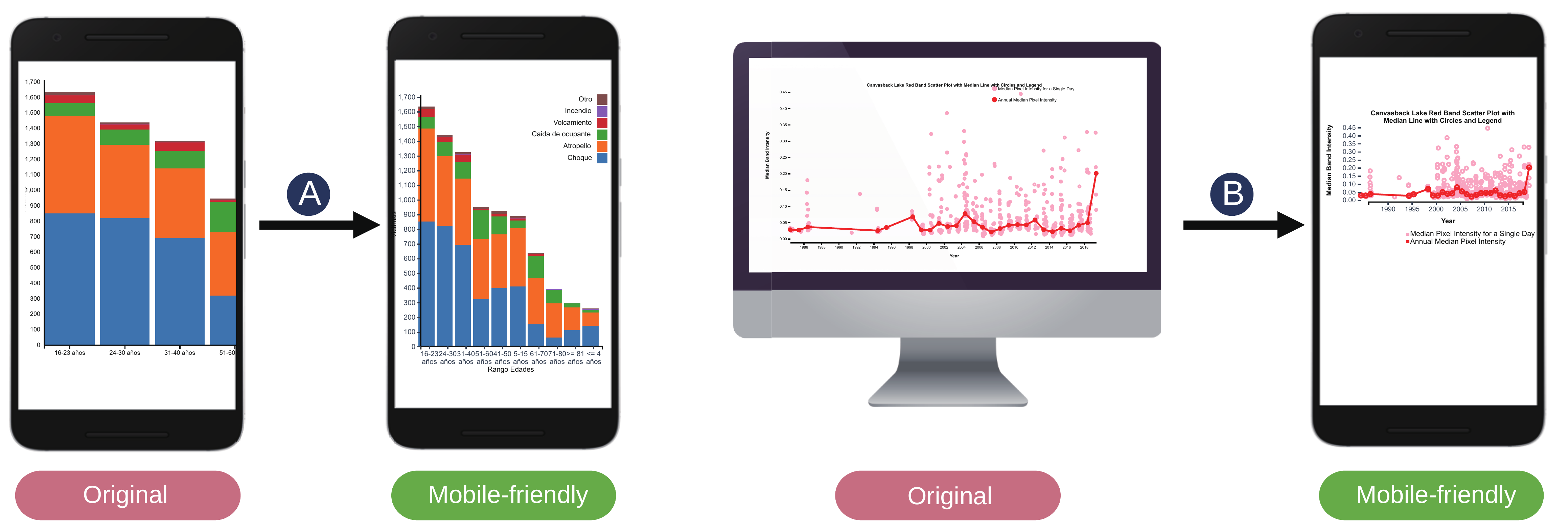}
  \caption{\mvf{} automatically transforms SVG-based visualizations from mobile or desktop versions into mobile-friendly designs with several modifications: (A) resize the view and modify axes; and (B) resize the view, reposition the legend, modify the title and axes.
  }
\label{fig:teaser}
}

\vgtcinsertpkg

\begin{document}



\firstsection{Introduction}
\maketitle

The last decade has seen an explosive growth of smartphone usage: statistics show that mobile devices have been used more than traditional desktops for web access globally since 2016~\cite{gibbs:2016, Enge:2019}. 
It is, therefore, becoming increasingly important to develop mobile-friendly websites that are readable and usable on mobile devices. 
We see efforts to promote mobile-friendly websites from both industry and research communities. 
For instance, Microsoft \cite{BingTool} and Google \cite{GoogleTool} have developed tools to test mobile friendliness, and favor mobile-friendly websites for their search results, which contribute to a trend towards mobile-first design. 
In addition, there are commercial services \cite{Mobify} and research efforts \cite{mahajan2018automated, althomali2019automatic} to fix problems with mobile designs that could cause a frustrating experience.
These efforts are typically focused on the design and layout of websites for mobile devices and do not address the specific challenges of mobile visualization design. As a result, a considerable number of visualizations on the web suffer from readability and usability issues, such as visual clutter, overlapping or tiny text, and overflowing content\cite{Brych2018, Young2019}. 

Despite the increasing acknowledgment of the opportunities and importance of mobile data visualizations~\cite{roberts2014visualization,lee2018data, choe2019mobile,lee2020reaching}, little work has attempted to investigate and fix the problems with mobile web-based visualizations. 
From a theoretical aspect, we lack empirical studies to understand the types of problems that occur in mobile visualizations. 
Recent work \cite{mahajan2018automated} has identified common types of mobile-related problems for general websites, but they are not readily applicable to the visualization context.
From an applied perspective, existing approaches are limited in helping practitioners detect and repair problems with  visualizations on mobile devices.
For example, the mobile friendliness test tools mentioned earlier~\cite{BingTool,GoogleTool} do not properly handle SVG-based visualizations. 
Practitioners who carefully craft custom designs, therefore, have to manually test and verify their visualizations on different screen sizes, which is tedious and time-consuming. 
After detecting problems, practitioners often find it difficult to repair them~\cite{Brych:2018, Bremer:2019}.
Practitioners typically need to adjust multiple SVG elements and CSS style properties simultaneously, while ensuring that those adjustments do not introduce any new problems~\cite{mahajan2018automated}. 
Automated tools that help tailor visualizations for mobile devices are one way to address these challenges. 

\begin{figure}[]
    \setlength{\abovecaptionskip}{5pt}
    \setlength{\belowcaptionskip}{-10pt}
	\centering
	\includegraphics[width=1\columnwidth]{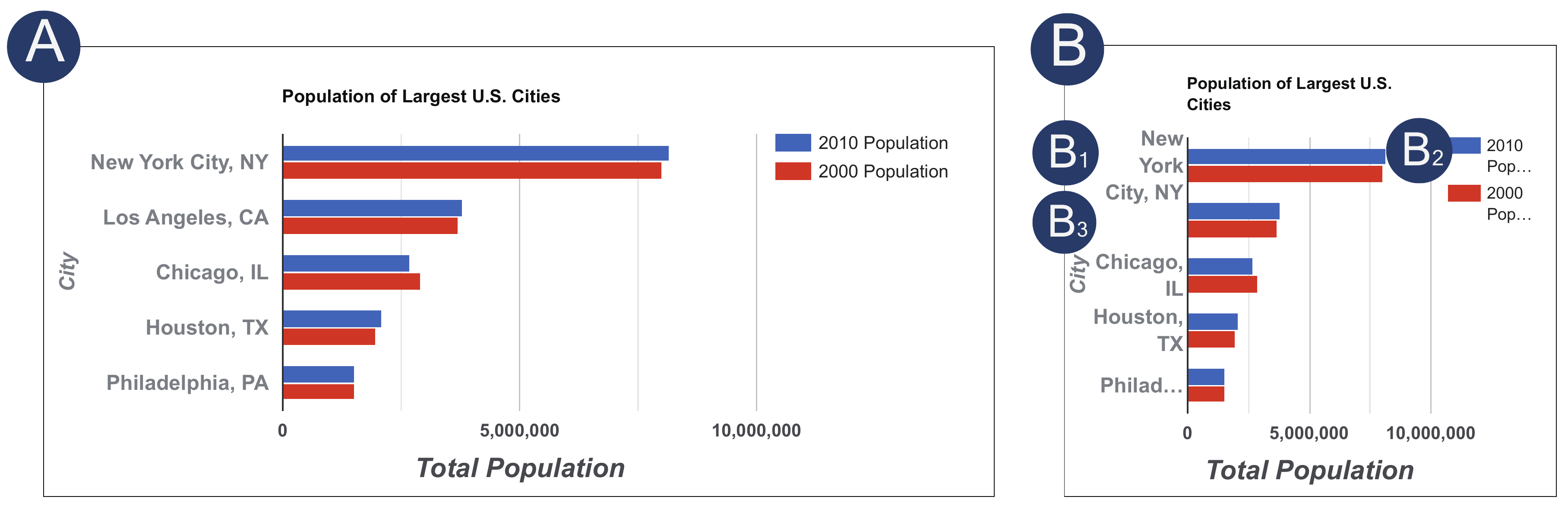}
	\caption{A motivating example of rule-based techniques for responsive visualizations\cite{motivateExample:20}: (A) The desktop version; (B) To fit visualizations on mobile devices, Google Charts employs rules including word-wrapping if not breaking words (B\textsubscript{1}), using ellipses else-wise (B\textsubscript{2}), and removing labels if overlapping with others (B\textsubscript{3}).}
	\label{fig:movEx}
\end{figure}

\begin{rev}
Existing automated solutions \cite{Gal:2017, Teal:2019, motivateExample:20} usually adjust visualizations by rule-based methods, \textit{e.g.}, word-wrapping if text overflows.
Their decision rules are interpretable and almost operable in their built-in visualization options.
However, such rules are often deterministic that could result into sub-optimal results in real-world scenarios.
Fig. \ref{fig:movEx} illustrates a real-world example by Google Charts \cite{motivateExample:20} where it adjusts texts according to a set of rules.
Those adjustments lead to readability issues as some text components (\textit{i.e.,} Fig. \ref{fig:movEx} B\textsubscript{2}, B\textsubscript{3}) become invisible and unreadable.
Although it is possible to add new rules to handle the situation, such rule-based methods face challenges such as large manual efforts and the combinatorial explosion of possible conditions \cite{saket2018beyond}.
It remains challenging and time-consuming to design rules for automatic responsive visualizations that scale well to the real-world diversity.
\end{rev}



\begin{figure*}[t]
    \setlength{\abovecaptionskip}{5pt}
    \setlength{\belowcaptionskip}{-10pt}	\centering
	\includegraphics[width=0.9\textwidth]{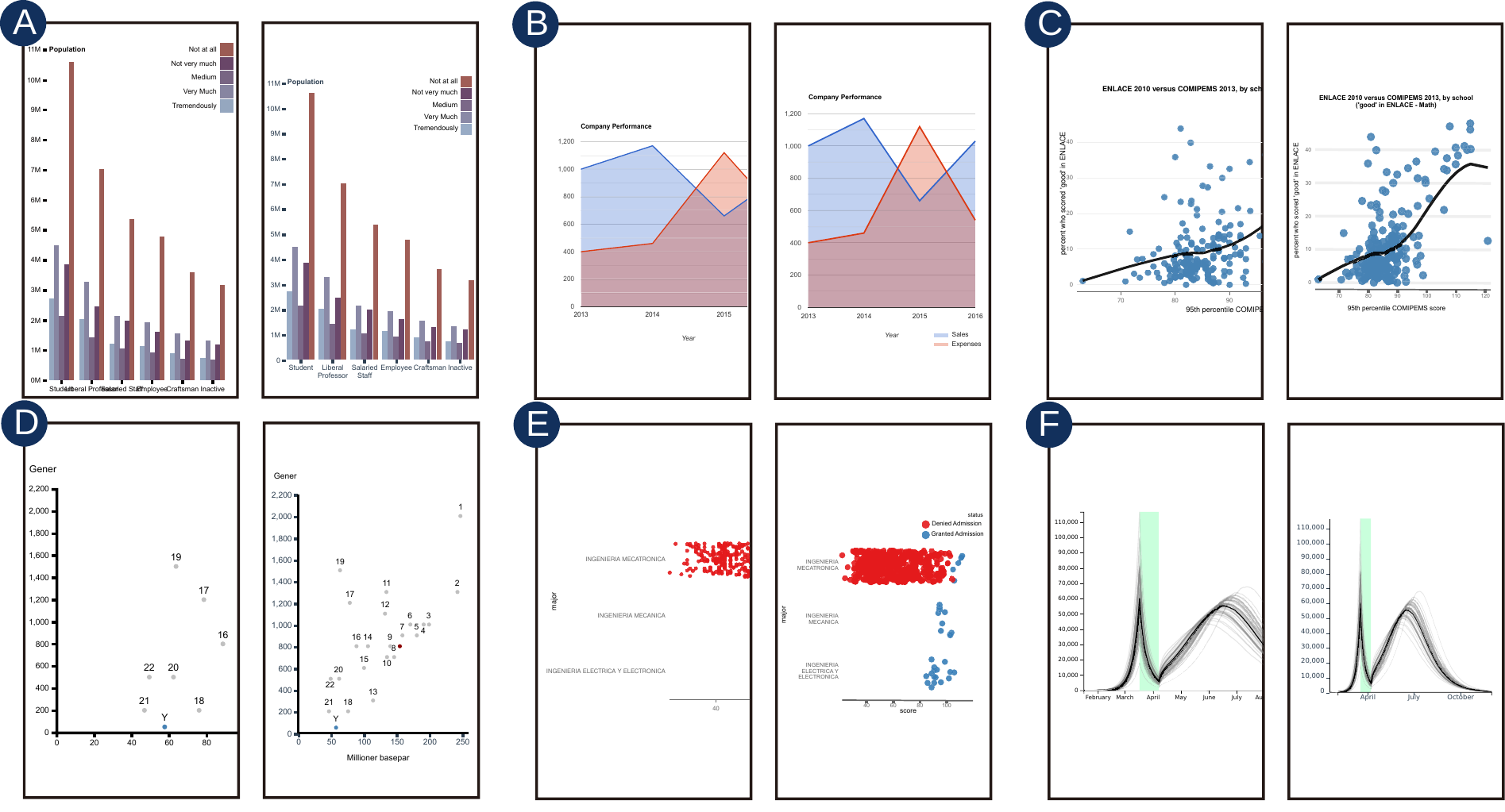}
	\caption{\mvf{} generates more mobile-friendly visualizations of various chart type while preserving original information. Visualization origins: (A) \cite{exg1}; (B)~\cite{exg2}; (C)~\cite{exg3}; (D)~\cite{exg4}; (E)~\cite{exg5}; and (F)~\cite{exg6}.
	Modifications: (A) modify axes; (B) resize the view, reposition charts and legends; (C) resize the view and modify the title; (D) resize the view, reposition charts and labels, modify axes; and (E)(F) resize the view and modify axes.}
	\label{fig:gallery}
\end{figure*}

\begin{rev}
In this paper, we present \mvf{}, an interpretable reinforcement-learning-based approach that automatically learns and applies decision rules for generating mobile-friendly visualizations.
We focus on SVG-based, single-view Cartesian visualizations with linear or discrete scales, which are found common on the web \cite{battle2018beagle}. 
To motivate the design of \mvf{}, we collected 374 web visualizations and categorized problems we saw when displaying these visualizations on mobile devices into five types of common issues.
\end{rev}
To optimize visualizations, 
\mvf{} first deconstructs charts into a declarative format which not only captures the underlying data encoding but also efficiently reduces the difficult, multi-criteria optimization problem.
\mvf{} then utilizes a novel greedy heuristic based on Policy Gradient methods that solves the reduced optimization problem.
\begin{rev}
Quantitative evaluation on a real-world dataset shows that \mvf{} successfully solve 89\% of visualizations with mobile-friendly problems in reasonable time: Figs.~\ref{fig:teaser} and ~\ref{fig:gallery} show several mobile-friendly visualizations that were automatically generated with \mvf{}.
\end{rev}
Further evaluation of \mvf{} on a different dataset demonstrates the generalizability of the learned model.

In summary, the primary contributions of this paper are:

\begin{compactitem}
    \item A categorization of five common issues with mobile visualizations derived from 374 web-based visualizations.
    \item The design and implementation of \mvf{}, which automatically converts SVG-based visualizations into mobile-friendly designs. \mvf{} takes an explainable machine learning approach to optimize the resolution of four problems across a large set of mobile visualizations.
    \item A set of quantitative evaluations that demonstrate the effectiveness, generalizability, and explainability of \mvf{}.
\end{compactitem}


\section{Related Work}
This paper draws upon prior work at the intersection of mobile web and mobile visualization, machine understanding of visualization, as well as automated visualization design.

\subsection{Mobile and Mobile Web Visualization}
There is a growing body of work examining how to adopt desktop web content to mobiles.
Typical approaches include \textit{Responsive Web Design} \cite{mullins2015responsive} that dynamically responds to size changes of the browser window using fluid grids and CSS media queries, as well as \textit{Adaptive Web Design} \cite{gustafson2015adaptive} that detects screen size and selects an appropriate design from multiple alternatives. 
However, both approaches introduce considerable development and testing costs, as developers must verify web-page appearance through trial-and-error.
As such, much research in the software community has studied how to automatically detect \cite{walsh2015automatic, walsh2017automated, althomali2019automatic} and repair \cite{mahajan2018automated, mahajan2017automated} mobile-friendly issues.
In contrast to our work, none of those approaches is targeted at visualizations.
In particular, existing techniques do not consider layout constraints in visualizations, and would potentially break the visual encoding and data binding.
Moreover, they do not support SVG which is the basis of many web-based visualization. SVG is a difficult target because it has its own set of elements, attributes, and properties that make it more complex than HTML alone.

In order to approach fixing problems with mobile visualizations we considered past design guidelines from the Visualization community. 
Already 14 years ago, Chittaro \cite{chittaro2006visualizing} argued that the different characteristics of mobile visualizations present new research challenges.  
Since then, research has proposed and evaluated mobile encodings for a wide range of data-types such as temporal data \cite{kay2016ish, chen2017visualizing, brehmer2018visualizing}, spatial-temporal data \cite{kim2007visual}, or small multiples \cite{brehmer2019comparative}. 
In contrast to this work on the development and study of dedicated mobile encodings is research that looked at how to adapt larger visualizations to smaller screen. 
Hoffswell \textit{et al.}~\cite{hoffswelltechniques} recently conducted a survey of existing practices for responsive visualization design and subsequently developed a tool to help people manually edit visualizations for different screen sizes.
On the commercial side, software like Power BI \cite{Gal:2017} and Tableau \cite{Teal:2019} also offered support for responsive layout but they do not proactively detect and diagnose potential problems.
Our work adds to this stream of research by proposing novel approaches that automatically detect and repair issues in mobile visualizations.

\subsection{Machine Understanding of Visualization}
To detect and repair problems present in mobile visualizations, our work takes inspiration from past work on the automatic extraction and manipulation of information in visualizations.
In a broader sense, recent research has been devoted to enabling machines to understand data visualizations from different perspectives. 
A majority of work investigates ways to retrieve data from charts \cite{choi2019visualizing, cliche2017scatteract, al2015automatic, poco2017reverse,jung2017chartsense,savva2011revision}, while other research attempts to retrieve color mappings \cite{poco2017extracting} and visual importance \cite{bylinskii2017learning}. 
Furthermore, researchers have developed and evaluated deep neural networks that reason about data visualizations including performing graphical perception tasks \cite{haehn2018evaluating} and visual question answering tasks \cite{kafle2018dvqa}.

The work more closely related to ours is
Battle \textit{el al.}'s \cite{battle2018beagle} Beagle system, which analyzes general SVG-based visualizations and automatically classifies them by type. 
The approach is similar to ours in that it also targets the general SVG-based visualizations we focus. 
We, however, take a different focus on adjusting SVG attributes related to layouts and visual styles to alleviate mobile-friendly issues while preserving the visual encoding.
As such, related to ours are Harper and Agrawala's methods for re-styling visualizations \cite{harper2014deconstructing} and for generating reusable templates \cite{harper2017converting} as well as following work \cite{hoque2019searching} on how to infer the visual style and structure from visualization collections. 
However, they do not specifically address the relationships of visual styles among elements in visualizations, \textit{e.g.}, the layout relationship between text labels and corresponding marks. 
Our work contributes to this space by studying how to model and deconstruct such relationships from SVG-based visualizations. 


\subsection{Automated Visualization Design}

Generating precise and elegant data visualizations is considered difficult even for experts \cite{qin2019making}.
\begin{rev}
Several automated visualization design tools have been proposed to ease this process by \textbf{rule-based} or \textbf{model-based} systems.
Rule-based systems typically introduce a set of heuristic rules to recommend visual encodings \cite{mackinlay2007show,wongsuphasawat2017voyager} or generate layouts \cite{ren2018charticulator}.
They have proven effective since their rules span a rich range of carefully curated design constraints considering data types and encoding channels.
Nevertheless, it requires system designers to apply prior domain knowledge from empirical studies to manually construct rules and curate a rule set \cite{saket2018beyond}.
Therefore, recent research starts shifting to model-based systems such as Data2Vis \cite{dibia2019data2vis} and VizML \cite{hu2019vizml}, which recommend design choices that are learned from a large corpus through machine-learning models.
However, despite promising results, their models have not proven superior.
In addition, the comprehensive data collection and labelling process could be expensive.

Different from the above work that recommends visual design given data, we study how to automatically adapt the layouts of existing visualizations to mobile screens.
Our method is inspired by the recent success of \textbf{hybrid} systems that augments rules with machine learning models \cite{moritz2018formalizing, saket2018beyond}.
In particular, we proposes a novel explainable reinforcement learning framework that automatically learns and executes human-interpretable rules for adjusting layout parameters to improve mobile-friendliness.
Compared with existing rule-based responsive visualization techniques (e.g. \cite{Gal:2017, Teal:2019}) that are usually deterministic, our framework could learn stochastic decision rules (policies) that allows generating more flexible solutions to varying real-world scenarios.
Besides, our framework embraces algorithmic explainability and transparency which helps model developers debug mistakes in cost functions, and reason about the quality of the learned model. 


\end{rev}



\section{Mobile-Friendly Issues in Web Visualizations}
\label{preStudy}

To gain an insight into mobile-friendly issues in web visualizations, we collected and analyzed SVG-based visualizations on the web. Our focus on web-based visualization is motivated by the fact that visualizations are often consumed on mobile devices, custom-designed, and thus difficult to adjust for all viewing scenarios. 
We note that we focus on layout-related readability issues of SVG-based visualizations and do not consider interaction problems.

We developed a web crawler to collect SVG-based visualizations following Hoque and Agrawala's approach \cite{hoque2019searching}. 
As their results are mainly from the \textit{bl.ock.org} domain, we extended their seeding pages with other visualization portals, such as Google Charts.
In addition, we randomly visited the hyperlinks in the queue 
to increase the diversity.
We used the Device Mode by Chrome DevTools to crawl visualizations rendered on an iPhone X screen.
We also crawled the desktop version to help us reason about if the creators had attempted to adapt visualizations to mobile screens or simply scaled them down.
At the end, we obtained 374 visualization examples from 103 domains.

Two authors of this paper manually inspected all mobile visualizations and coded problems that hurt the appearance and readability of the mobile visualizations.
The coding schemes were originally based on existing literature about mobile-friendly problems in general web content \cite{GoogleTool, BingTool}, such as small font size and wrong viewport.
Throughout the coding, we iteratively updated the coding schemes and re-coded samples when necessary.
In the following text, we describe the most common sources of problems we found in detail.

\begin{figure*}[!ht]
    \setlength{\belowcaptionskip}{-10pt}
    \setlength{\abovecaptionskip}{5pt}
	\centering
	\includegraphics[width=2\columnwidth]{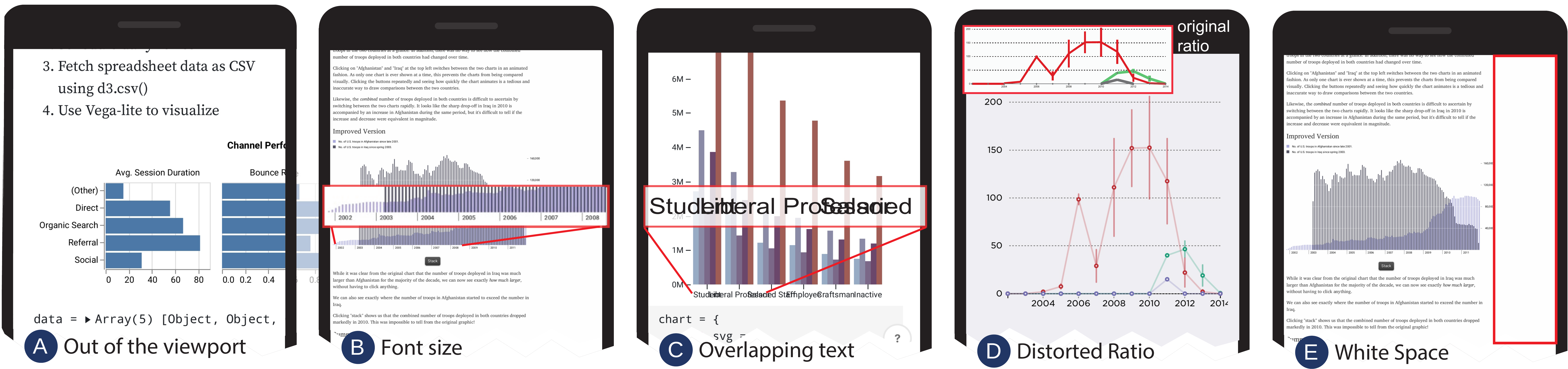}
	\caption{We identified five common mobile-friendly issues in SVG-based visualizations on mobile devices: (A) A chart is out of the viewport that requires users to scroll horizontally, which can easily get overlooked; (B) The font size is too small and unreadable; (C) The text labels in the axis suffer from visual clutter; (D) The height/width ratio differs from that of the desktop version, which could lead to perceptional bias; and (E) The visualization has much empty space on its left, which could be used more effectively.}
	\label{fig:mobFriendProblems}
\end{figure*}

\subsection{Mobile-friendly issues}
We identify five common issues that impair the mobile-friendliness of visualizations (Fig.~\ref{fig:mobFriendProblems}).
\begin{rev}
We discuss them together with the contributing inappropriate changes between desktop and mobile versions.
\end{rev}

\bpstart{1) Out of the viewport}
Out of 374 visualizations, 122 (32.6\%) had problems related to content being placed outside of the screen. This problem can occur when absolute or miscalculated values of SVG properties lead to display coordinates outside of the current viewpoint. This problem forces viewers to scroll horizontally to view the whole content, resulting in a poor user experience \cite{LePage:2019}.


\bpstart{2) Unreadable font size} A large number (118, 31.5\%) of visualizations included font sizes that were hardly readable. 
This problem occurs when programmers only resize visualizations to fit the current screen, making visualizations fully visible but making content less legible.

\bpstart{3) Cluttered text} About 16.0\% of visualizations (60) contained overlapping and cluttered text elements. 
This problem is partly due to the absence of intrinsic mechanisms for preventing overlap in SVG elements or no implemented ways to avoid label overlap. 


\bpstart{4) Distorted layout} For 85 (22.7\%) web-based visualizations the layout was artificially stretched. This problem occurs because mobile devices are predominantly held in portrait orientation even when browsing multimedia content \cite{Podger:2019}, while desktop web browsing is more typically in landscape mode.
To address this difference, web programmers often adjust web content to the screen's width, letting content spread out vertically. 
However, such practices, when applied in the SVG context, can render visualizations in a distorted aspect ratio that potentially causes unintended bias in visual perception \cite{talbot2012empirical}.

\bpstart{5) Unwanted white space} The fifth most common (21, 5.6\%) cause of problems in the SVG-based visualizations we coded was excess white space, leading to non-optimal space usage and potentially unreadable content.
This problem is often due to the use of fixed-width layout attributes such as padding and margins.

\begin{rev}
We found that 142 (37.9\%) visualizations exhibited no changes between desktop and mobile designs, while only 98 (26.2\%) visualizations exhibited none of the above five issues.
This indicates that many visualization creators might neglect responsive design.
Besides, we observed entanglement among issues. 
About 36.6\% (101) out of the 276 non-mobile-friendly visualizations contained more than one issue.
\end{rev}

Beyond these five common issues, we also found several rare cases pertained to mobile-friendliness.
A few visualizations embed third-party icons or images that do not automatically scale well to the mobile screen.
Besides, touch elements (\textit{e.g.}, buttons) can be too close to each other that users might have difficulties tapping desired ones. 

\subsection{Discussion}

Our analysis shows several visualization-specific problems for mobile content compared to those of general web content~\cite{mahajan2018automated}.
While content sizing and viewport configuration are commonplace for both visualizations and general web content, research by Mahajan \textit{et al.}\ \cite{mahajan2018automated} does not discuss our last three issues (\textit{i.\,e.}, cluttered text, distorted layout, and unwanted white space).
%
 This underscores the potential for developing an automatic, visualization-specific approach.

It is challenging to address all five issues we uncovered simultaneously, since they are highly inter-dependent. For example, increasing the font size to make text legible might lead to overlapping text making it illegible.
Therefore, designers must fix related problems in a trial-and-error process, often leaving some problems not optimally solved.
Furthermore, designers might fail to anticipate data changes that could distort the layout \cite{walny2019data}.
For example, Fig.\ \ref{fig:mobFriendProblems} (C)~\footnote{https://observablehq.com/@2shabby/grouped-bar-chart} forks a template~\footnote{https://observablehq.com/@d3/grouped-bar-chart} of a grouped bar chart.
However, its labels are considerably longer than that of the template and the mobile version is compromised.


\section{\mvf{} - Overview}
\label{mvf}

\mvf{} automatically generates mobile-friendly designs for SVG-based visualizations.
It currently addresses four common types of problems introduced in \autoref{preStudy}: \emph{content sizing for viewport, font sizing, text overlap,} and \emph{white space}. 
\begin{rev}
We leave the last one, \emph{distorted ratio}, to future work because it requires perceptual guidelines for a large number of visualizations for which they are not yet clearly defined.
Specifically, due to the different screen size between desktops and mobiles,
resizing is the most common compromise solution for responsive visualization design~\cite{hoffswelltechniques} that usually distorts the aspect ratio.
It, however, remains unclear to what extend that such distortions influence perceptions.
\end{rev}

There are some straightforward repairs for the four issues in graphics design~\cite{o2014learning}, \textit{e.g.}, moving text or graphical marks elsewhere to prevent overlap.
However, naive movements of elements can easily violate data representations where positions are mapped to data.
Therefore, the challenge of generating a reasonable repair involves two objectives--addressing multi-criteria mobile-friendly issues and strictly maintaining the underlying visual encoding.


\mvf{} transforms visual encoding to a mathematical form. 
It defines a visualization as a set of visual \textit{elements} $(e \in \mathcal{E})$, including text and graphical marks. 
Each element is described by a visual \textit{encoding} which specifies \textit{values} $(v_{e, p} \in \mathbb{R})$ for visual \textit{properties} $(p \in \mathcal{P})$, such as positions and sizes in SVG attributes and CSS styles. 
We use a simplified notation to consider all values in the real domain, assuming that categorical attributes can be expressed by enumerations.
Let $\mathcal{C}$ denote all valid element-property pairs, where $\mathcal{C} \subseteq	{\mathcal{E} \times \mathcal{P}}$.
The visualization is thereby expressed as a vector containing values for all those pairs, namely $\chi \in \mathbb{R}^{|\mathcal{C}|}$.
\mvf{} quantifies the mobile-friendly issues with a multi-criteria cost function, denoted \((J)\). 
The objective is to determine a set of \textit{patches}, denoted $\chi^*$, that minimizes \(J\):

\begin{figure}[]
    \setlength{\abovecaptionskip}{5pt}
    \setlength{\belowcaptionskip}{-15pt}
	\centering
	\includegraphics[width=0.9\columnwidth]{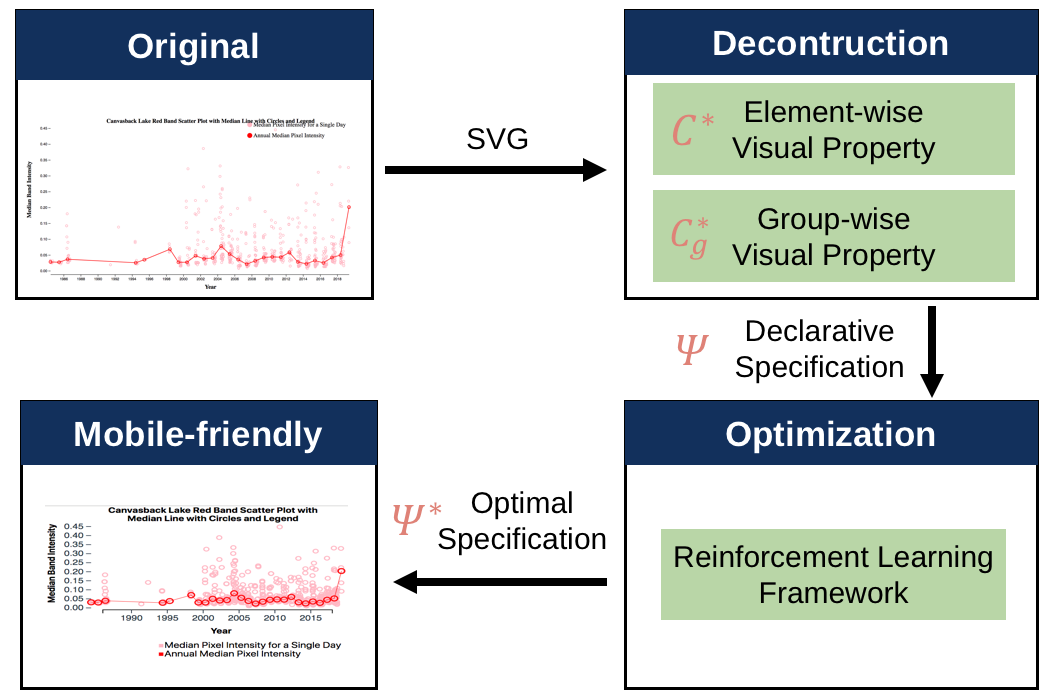}
	\caption{Given only an SVG as input, \mvf{} generates mobile-friendly designs in two steps: deconstruction and optimization. }
	\label{fig:overview}
\end{figure}

\begin{equation}
\label{optFunc}
\chi^* := \operatorname*{argmin}_{\chi \in \mathbb{R}^{|\mathcal{C}|}}  J(\chi)
\end{equation}

Our approach for solving the above multi-criteria, high-dimensional optimization problem consists of two phases, deconstruction and optimization, as shown in \autoref{fig:overview}. 
The input to \mvf{} is an SVG file containing a visualization to be rendered on mobile devices. 
The deconstruction phase (\autoref{mvf:decode}) decodes the visualization to extract the data and encoding, which are described in a declarative format. 
The output is $\psi$, the parameters used in the declarative descriptions of visualizations, which are reduced from $\chi$ in the convenience of solving effectively.
The optimization phase (\autoref{mvf:framework}) proposes a novel explainable reinforcement learning framework that solves (\autoref{optFunc}) and generates optimal $\chi^*$, as well as the corresponding visualization.

The term ``optimal'' here refers to a reasonable solution that minimizes built-in costs and thus improves mobile-friendliness for a particular visualization. 
It does not mean ``optimal'' for any specific set of requirements (\textit{e.g.,} data, context) of a human designer.
A designer, instead could view different results from our tool and select from or refine multiple  generated designs until they find the ``best'' designs for their own tasks and additional considerations \cite{moritz2018formalizing}.

\section{Deconstruction}
\label{mvf:decode}

\begin{figure*}[]
    \setlength{\abovecaptionskip}{5pt}
    \setlength{\belowcaptionskip}{-10pt}
	\centering
	\includegraphics[width=1.9\columnwidth]{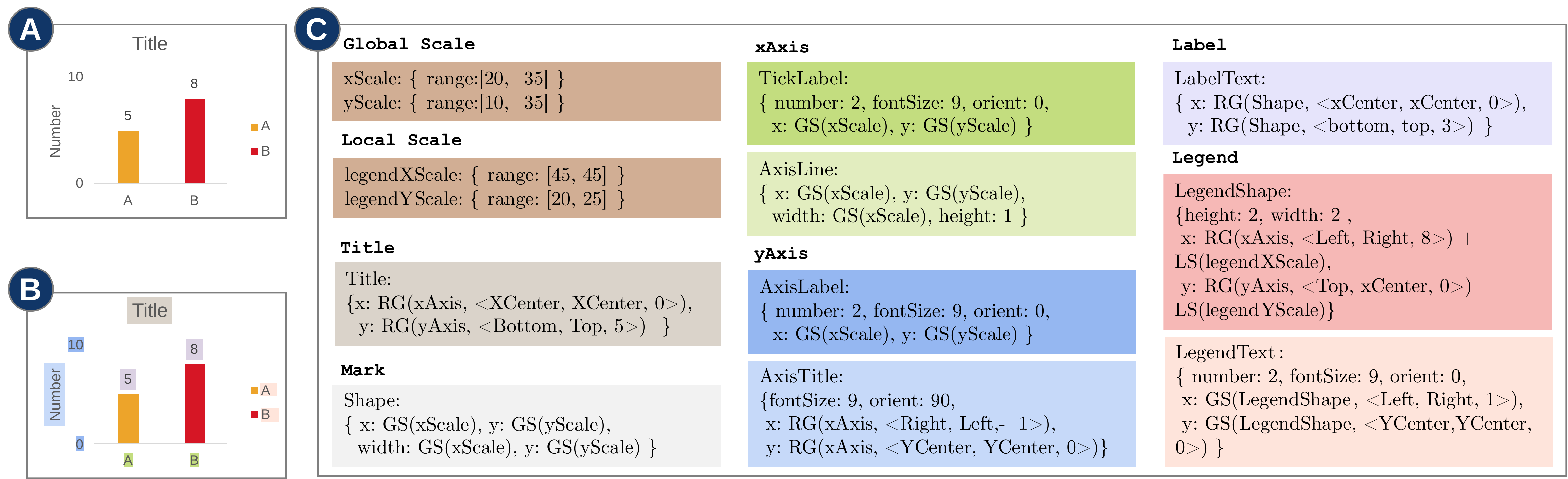}
	\caption{Decoding phase: (A) The input is the original chart; (B) The first step identifies visual groups - a set of elements with the same encoding; (C) The second step determines the group-wise encoding in a declarative format. All the parameters constitute the output $\psi$.}
	\label{fig:decoding}
\end{figure*}

The deconstruction phase decodes the SVG to identify $\mathcal{C}^*$ -- sets of SVG elements with visual properties that are subject to adjustments to improve the mobile-friendliness,
as well as $\psi$ -- a declarative description of visualizations that facilitate the computation.
The general intuition of this decision is two-fold: (1) most encoding properties (\textit{e.g.}, color, border) should remain faithful to the original chart; and (2) each mobile-friendly issue typically maps to a small set of properties.
This reduces the solution space of $\chi$ in \autoref{optFunc} from $\mathbb{R}^{|\mathcal{C}|}$ to $\mathbb{R}^{|\mathcal{C}^*|}$.
However, this phase needs to consider constraints for the change of properties that are related to the underlying data binding and visual consistency.

To consider those constraints, \mvf{} introduces visual \textit{groups} $(g \in \mathcal{G})$ --  sets of elements described by the same set of encodings, \textit{e.g.}, the tick-labels in an axis form a visual group.
Following the same notation used for element-wise encoding ($\mathcal{C}$), group-wise encodings can be expressed as $\mathcal{C}_g^* \subseteq \mathcal{C}_g \subseteq	{\mathcal{G} \times \mathcal{P}}$, and corresponding values $\chi_g \in \mathbb{R}^{|\mathcal{C}_g^*|}$.
Thus, the solution space is further reduced to $\mathbb{R}^{|\mathcal{C}_g^*|}$. 
\mvf{} determines $\mathcal{C}_g^*$ by deconstructing the visual encoding through two steps: (1) generating visual groups and their intra-group encoding and (2) determining \textit{layout dependencies} -- the inter-group layout relationships.

The first step builds on Hoque and Agrawala's \cite{hoque2019searching} method for recovering data, marks, and encoding from a D3 chart.
Due to the complexity and diversity of data visualizations, \mvf{} currently focuses on single Cartesian visualizations with linear or discrete scales.
\mvf{} extends their method to support non-D3 charts by dismissing D3's specification for the SVG tree structure.
For instance, D3\footnote{https://github.com/d3/d3-axis} utilizes a  \mintinline{html}{<g>} template for rendering axes.
In addition to searching and traversing such \mintinline{html}{<g>} nodes, \mvf{} also uses a linear scan algorithm to search axis candidates with aligned tick-labels and ticks. 
The result of this step is a set of visual groups, as well as coordinate scales.
\autoref{fig:decoding} (B) shows each visual group annotated by the same background color, except for the line in the x-Axis and the bars which already has their own background color.

The second step aims to identify the layout dependency of the visual group. 
Similar to Vega's specification\cite{satyanarayan2016vega}, the layout of a group depends on either the coordinate scale or another \textit{anchoring group}.
The latter case is referred to as \textit{reactive geometry} which is particularly common for describing the layout of text labels.
\mvf{} describes reactive geometry by a tuple \(\langle p, p_a, o \rangle\), where \(p\) and \(p_a\) are the \textit{anchoring position} for the group and anchoring group respectively, and \(o\) is the offset value in corresponding direction.
Possible anchoring positions include \emph{Left, X-Center, Right, Top, Y-Center,} and \emph{Bottom.}
For instance, the labels in \autoref{fig:decoding} (A) are horizontally aligned center to their corresponding bars, which is described as $\langle$X-Center, X-Center, 0$\rangle$.

The resulting specification of visualizations is similar with Vega \cite{satyanarayan2015reactive}, a declarative format based on grammar-based specifications.
\begin{rev}
We choose a subset of Vega specifications related to layouts.
As shown in Table \ref{table:specification}, \mvf{} includes five classes (Title, Axis, Legend, Mark, and Label) and 10 visual groups, which are basic structural elements of visualizations.
Different from Vega where labels are included in marks, we consider labels as a separate class since they have unique encoding properties such as font-sizes.
\end{rev}
For each group, \mvf{} identifies a set of encoding properties which are subject to adjustments to improve mobile-friendliness.
For instance, adjustable properties for \tGroup{TitleText} includes \formatHTML{<fontSize>}, \formatHTML{<orient>}, and layout properties such as \formatHTML{<x>} and \formatHTML{<y>}.
All those group-property pairs form $\mathcal{C}_g^*$.

\begin{table}[t]
\begin{tabular}{p{0.75cm} p{1.6 cm} p{5.2cm}}
\toprule
\textbf{Class}                 & \textbf{Group}        & \textbf{Adjustable Encoding Properties}                                            \\ \hline
\tClass{Title}                 & \tGroup{TitleText}    & \tLdep{RG}{AxisLine}, \tEncoding{<fontSize>}\tEncoding{<orient>}             \\ \hline
\multirow[t]{4}{*}{\tClass{Axis}} & \tGroup{AxisTick}     & \tLdep{GS}{}, \tEncoding{<number>}                    \\ 
                               & \tGroup{AxisLabel}    & \tLdep{RG}{AxisTick}, \tEncoding{<number>} \tEncoding{<fontSize>} \tEncoding{<orient>}            \\
                               & \tGroup{AxisLine} & \tLdep{GS}{}, \tEncoding{<height>} \tEncoding{<width>}       \\
                               & \tGroup{AxisTitle} & \tLdep{RG}{AxisLine}, \tEncoding{<fontSize>}\tEncoding{<orient>}        \\
                               & \tGroup{Grid}         & \tLdep{GS}{}, \tEncoding{<number>}                  \\ \hline
\multirow[t]{2}{*}{\tClass{Legend}} & \tGroup{LegendShape}  & \tLdep{RG}{AxisLine}, \tLdep{LS}{}, \tEncoding{<height>} \tEncoding{<width>} \tEncoding{<radius>} \\  
                               & \tGroup{LegendText}   &  \tLdep{RG}{LegendShape}, \tEncoding{<fontSize><orient>} \\ \hline
\tClass{Mark}                  & \tGroup{Shape}        & \tLdep{GS}{}               \\ \hline
\tClass{Label}                 & \tGroup{LabelText}    & \tLdep{RG}{Shape}, \tEncoding{<fontSize>}\tEncoding{<orient>}   \\ 

\bottomrule

\end{tabular}
\\[1pt]
\textbf{Notation}: Encoding includes \tLdeps{layout dependency} and \tEncoding{independent} \tEncoding{attributes}. The former includes \tLdeps{Global} \tLdeps{Scale (GS)}, \tLdeps{Local Scale (LS)}, or \tLdeps{Reactive Geometry (RG)}.
\\[1pt]
\vspace{-5px}
\caption {\mvf{} identifies 5 classes and 10 groups that effectively represents a data visualization. Each group has a set of adjustable encoding properties. All those group-property combinations yield $\mathcal{C}_g^*$.}
\label{table:specification}
\vspace{-15px}
\end{table}

\mvf{} classifies adjustable properties into independent and dependent attributes.
Independent attributes are expressed as \tEncoding{variables}, while dependent attributes are \tLdeps{function-like}.
\mvf{} currently only considers the dependency relationships of layout-related properties (\textit{e.g.}, x, y coordinates), including Global Scale (GS), Local Scale (LS), and Reactive Geometry (RG).
For example, the positions of marks depend on the global coordinate scales, while titles and legends are located based on the axes positions.
In addition, a legend also has its own local scale for placing its constituent shapes and text, and legend texts have a reactive geometry depending on corresponding legend shapes.
Fig.~\ref{fig:decoding} (C) shows an example. 

There exist alternative specifications for the aforementioned layout dependency, since the mapping $\chi_g^* \to \psi$ is potentially one-to-many.
The layouts of titles and legends could directly map to the global scale rather than using reactive geometry. 
The advantage of the latter is that it supports quick modifications of inter-group layouts through discrete operations, \textit{e.g.}, switching the anchor position from top to bottom to render titles underneath the chart, whereas the former would require updating the vertical position in a continuous space.
In \autoref{mvf:framework:algo} we discuss how such discretization provides conveniences for effectively solving the problem.

We note that \mvf{} does not consider potential dependencies of non-layout properties a designer might have chosen. For instance, there might exist a dependency between font-sizes of title and labels. Future work should address describing such dependencies.
By adopting the declarative specifications, \mvf{} maps the group-wise visual encoding  $\chi_g^* \in \mathbb{R}^{|\mathcal{C}_g^*|}$ to a parameter space ($\psi$).
For instance, a group-property pair (\tGroup{AxisTick}, \tLdeps{GS}) is mapped to declarative parameters (\formatHTML{scaleRangeMin, scaleRangeMax}).
Eventually, the optimization problem (\autoref{optFunc}) is reduced to 
\begin{equation}
\label{optFunc2}
\psi^* := \operatorname*{argmin}_{\psi}  J(\psi)
\end{equation}
\section{Optimization}
\label{mvf:framework}
\mvf{} proposes an explainable reinforcement learning framework for solving the optimization problem.
We present our design goals, the Markov Decision Process (MDP) model, and the heuristic.

\subsection{Design Goals}
\mvf{} is designed to improve common readability and aesthetics issues of web-based visualization while requiring minimal human intervention.
We set ourselves the following design goals that aim to 
increase practical applicability.

\bpstart{G1: Simulate a trial-and-error process for manual repair}
Manual creation of mobile-friendly visualizations is known to be an \textit{ad hoc.}, iterative process \cite{hoffswelltechniques}, involving the adjustment of visual encodings while ensuring that these adjustments do not impact other parts of the visualization.
\mvf{} aims to automate this process by mimicking human behavior.

\bpstart{G2: Ensure the transparency and explainability of the automation}
Algorithmic interpretability and transparency are increasingly important for automated systems.
By utilizing explainable approaches, \mvf{} aims not only to support us in understanding our model, but also to help us gain insights for designing mobile-friendly visualizations by summarizing what machines have learned.

\bpstart{G3: Remain as faithful as possible to the original visualization}
Visualizations are usually crafted with deliberate designs, which machines could fail to understand and preserve.
As an automated framework, \mvf{} strives to maintain the visual encoding and design.

\bpstart{G4: Support compatibility with other algorithms}
Mobile-friendly issues of visualizations are complex and even potentially ill-posed -- a one-size-fits-all solution may not exist, which does not require modifications by a human.
\begin{rev}
\mvf's goal is to alleviate such challenges by supporting existing algorithms for optimizing visualizations.
\end{rev}

\bpstart{G5: Execute in browser rendering time}
The last goal of \mvf{} concerns practical applicability -- the automatic process should terminate in approximately similar time to the browser rendering process to meet real-world performance needs.

\subsection{Explainable MDP Model}
\label{mvf:framework:model}
\mvf{} uses a reinforcement learning framework to solve the optimization problem, because reinforcement learning can theoretically mimic human behavior by learning from rewards \textbf{(G1)} \cite{peng2018deepmimic}.

\begin{figure}[t]
    \setlength{\abovecaptionskip}{0pt}
    \setlength{\belowcaptionskip}{-5pt}
	\centering
	\includegraphics[width=1\columnwidth]{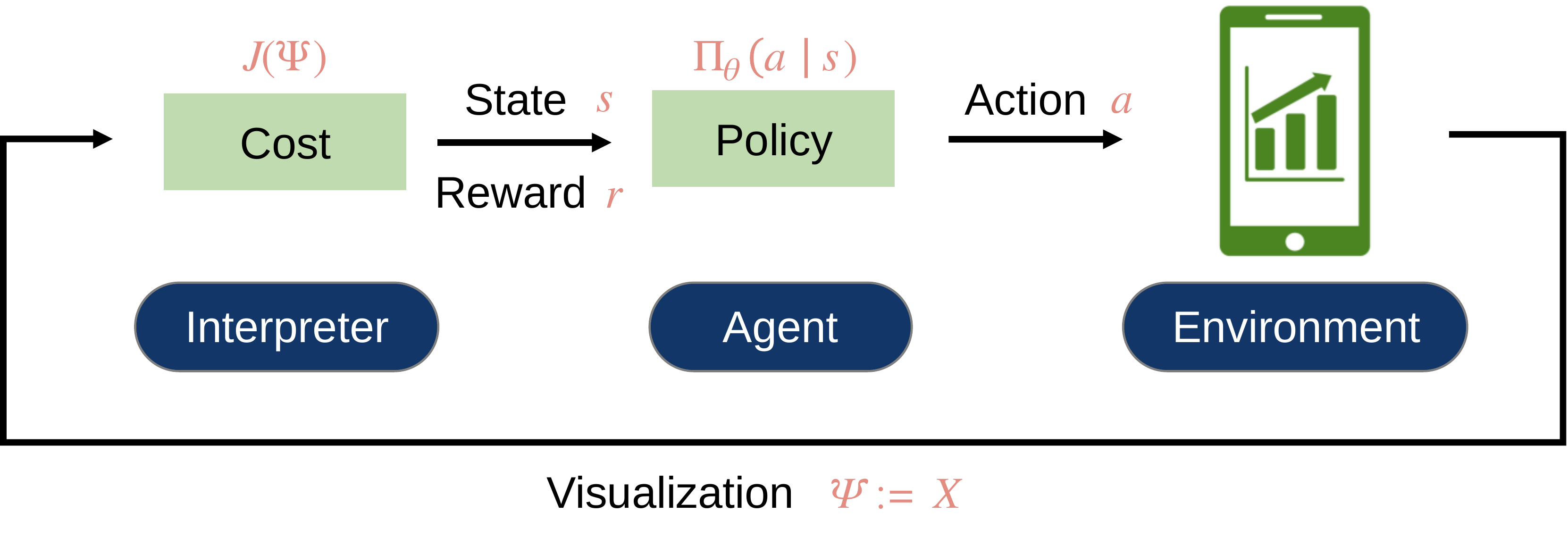}
	\caption{Illustration of the reinforcement learning framework.}
	\vspace{-1mm}
	\label{fig:RLframework}
    \vspace{2mm}
\end{figure}

A reinforcement framework is modelled as a Markov decision process (MDP), as illustrated in \autoref{fig:RLframework}. 
The \textit{environment} is the visualization specified by declarative parameter $\psi$.
An \textit{interpreter} calculates the cost $J(\psi)$ in respect to mobile-friendly issues.
The \textit{agent} observes the \textit{state} ($s \in \mathcal{S}$) and \textit{reward} ($r \in \mathbb{R}$), thereby taking an \textit{action} ($a \in \mathcal{A}$) to manipulate $\psi$ and consequently the environment.
The agent's action selection is based on the policy $\Pi(a | s)$ - the probability that the agent takes action $a$ when in state $s$.
Thus, the goal is to learn the optimal policy that maximizes the rewards and therefore solves \autoref{optFunc2} effectively.
In the following text, we explain the states, actions, and costs in detail.

\begin{table}[]
\begin{tabular}{p{0.6cm} p{3cm} p{4.2cm}}
\toprule
\textbf{Scope}         & \textbf{Mobile-friendly Issues} & \textbf{Notations}                                      \\ \hline
Global                 & Unwanted White Space           & LeftMargin, RightMargin, TopMargin                      \\ \hline
\multirow[t]{3}{*}{Local} & Out-of-viewport                 & LeftOutOfViewport, RightOutOfViewport, TopOutOfViewport \\
                       & Unreadable Fontsize             & FontSize                                                \\
                       & Overlapping Text                & OverlappingText         \\ 
                       \bottomrule                               
\end{tabular}
\\[3pt]
\caption {Summary of the notations of mobile-friendly issues.}
\label{table:issues}
\vspace{-15px}
\end{table}

\subsubsection{State}
\label{mvf:framework:model:state}
\mvf{} features explicit definitions of states for framework explainability (\textbf{G2}).
Specifically, states describe mobile-friendly issues that are observable by both humans and computers (\textit{i.e.}, the interpreter).
The notations of those issues, denoted $\mathcal{N}$,  are summarized in Table \ref{table:issues}.
\mvf{}  assumes that a visualization can be scrolled infinitely towards the bottom, and, thus the bottom orientation is not included in the out-of-viewport category.
\mvf{} classifies notations into global ($\mathcal{N}_G$) and local ($\mathcal{N}_L$).
The former applies to the global visualization, while the latter is specific to an individual visual element.
Consequently, the total number of possible issues in a visualization is $|\mathcal{N}_G| + |\mathcal{E}| \times |\mathcal{N}_L|$.
Consider that those issues could appear simultaneously which means that the total number of states becomes $2^{|\mathcal{N}_G| + |\mathcal{E}| \times |\mathcal{N}_L|}$ which renders the time complexity exponential.
In addition, as visualizations can vary from $|\mathcal{E}|$, the model is not generalizable across visualizations.
To alleviate those challenges (\textbf{G5}), \mvf{} proposes two strategies for State Aggregation -- a common technique to reduce the number of states \cite{sutton2018reinforcement}:

\bpstart{1) Leveraging domain knowledge} We take advantage of our domain knowledge of visualizations.
In particular, we aggregate visual elements into their corresponding classes (Title, Axis, Legend, Mark, and Label), as shown in  \autoref{table:specification}.
The resulting number of states is therefore reduced to  $2^{|\mathcal{N}_G| + 5 \times |\mathcal{N}_L|}$.
Furthermore, such state aggregation allows some generalizability 
as those classes are universal across visualizations, as demonstrated in \autoref{eva}.

\bpstart{2) Greedy aggregation} The result after the above step still has exponential complexity, which motivates us to adopt a greedy design -- an established method for approximation algorithms \cite{vazirani2013approximation}.
Instead of considering all mobile-friendly issues simultaneously, \mvf{} greedily selects only one as the current state and, upon solving it, moves to the next issue as another state.
This greedy strategy reduces the total number of states to $|\mathcal{N}_G| + 5 \times |\mathcal{N}_L|$, which becomes polynomial.

It should be noted that the optimization problem (\autoref{optFunc2}) is indeed a multi-objective reinforcement learning problem -- one that could require compromising solutions that balance different objectives (\textit{i.\,e.}, mobile-friendly issues) \cite{van2014multi}.
The above state aggregation naturally leads to an approximation algorithm, by greedily solving a single-objective reinforcement learning problem.
This inherits disadvantages of greedy algorithms which may make commitments to non-optimal solutions too early \cite{devore1996some}. 
We will discuss our solution to this challenge in \autoref{mvf:framework:algo}.

\subsubsection{Action}
\label{mvf:framework:model:action}
Actions in \mvf{} manipulate the data visualization by updating the parameter space $\psi$.
Most actions are defined by incremental or decremental operations due to two considerations:
first, such progressive changes are based on the original value (\textbf{G3}); 
and second, it allows us to discretize continuous action space, which could significantly improve the efficiency \cite{munos2002variable} (\textbf{G5}).

To be specific, inheriting from $\chi$, the parameter in $\psi$ (\textit{e.g.}, the scale range) are in continuous space $\mathbb{R}$.
Thus, the action space which manipulates those parameters is also continuous and infinite.
\mvf{} discretizes this infinite space through incremental or decremental operations, \textit{e.g.}, increase the min value of the scale range by a fixed number $\Delta$.
Users could specify the value of $\Delta$ (5px by default).

As shown in \autoref{fig:policy}, \mvf{} currently has 23 actions. 
Twenty-one actions are incremental or decremental operations for global scale (8), local scale and reactive geometry (8), font-size (2), tick number of axes (2), and anchoring position of reactive geometry (1).
Besides, it includes 2 actions which executes third-party algorithms for optimizing visualizations\footnote{\label{footnoteVegaLabel}https://github.com/vega/vega-label}\textsuperscript{,}\footnote{https://bl.ocks.org/mbostock/7555321}.
While it currently only utilizes two third-party algorithms, more can be added in the same manner (\textbf{G4}).

\subsubsection{Cost Function}
\label{mvf:framework:model:cost}
\mvf{} defines the cost function $J(\psi)$ in respect to each mobile-friendly issue.
The cost function is designed iteratively -- during development we found that a seemingly reasonable definition could lead to unexpected behavior of machines.
Thanks to the explainable nature of our framework, we were able to locate the root cause in cost functions and make corrections. Below, we introduce cost functions in turn.

\begin{figure}[!h]
    \setlength{\abovecaptionskip}{0pt}
    \setlength{\belowcaptionskip}{-5pt}
	\centering
	\includegraphics[width=1\columnwidth]{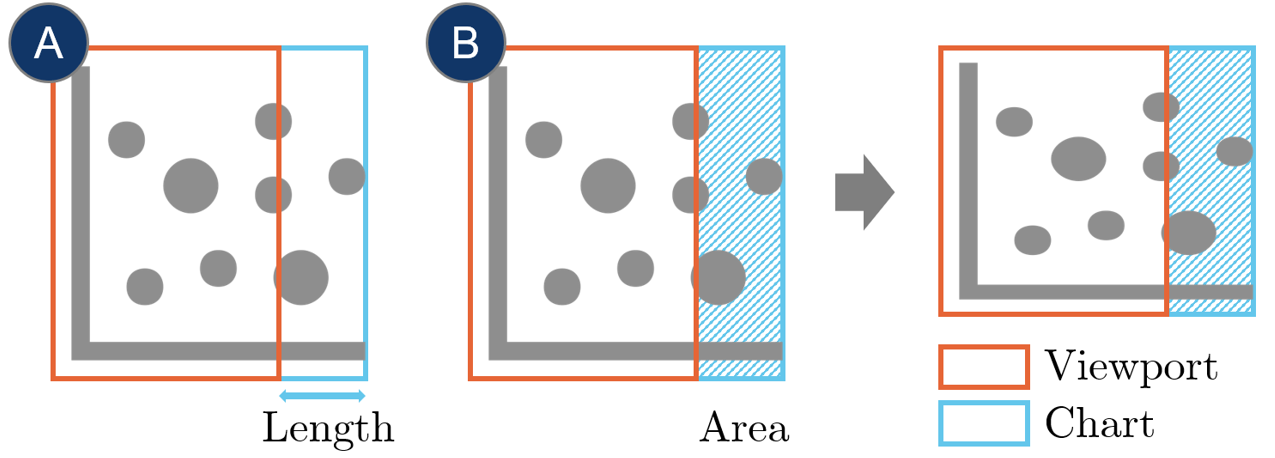}
	\caption{(A) The cost function of out-of-viewport is determined by the length exceeding it; (B) A seemingly reasonable alternative is the area. However, the agent can learn to compress the height to reduce the cost.}
	\label{fig:costFunOutView}
\end{figure}

\bpstart{Out of the viewport} The cost is defined by the length exceeding the viewport on the left, right, and top orientation (\autoref{fig:costFunOutView} (A)).
Initially we defined the cost by the area outside the viewport. 
However, the machine tended to reduce the visualization height when in the out-of-viewport state (\autoref{fig:costFunOutView} (B)).
This action reduced the costs but yielded no actual improvement. 
More reasonable costs include the length or the relative area, and we utilize the former which is validate after experiments.

\bpstart{Unwanted white space} Similar to the above, the cost is determined by the length of margins that exceed the threshold.
The thresholds for amount of whitespace can be determined by users.

\bpstart{Unreadable font-size} The cost is calculated by:
\begin{equation}
\setlength{\belowdisplayskip}{3pt}
\setlength{\abovedisplayskip}{3pt}
\label{optFuncFontSize}
\frac {1}{n}\sum _{i=1}^{n}f^+({s_i - \tau})
\end{equation}
where $s_i$ is the font-size of the $i$-th text, and $\tau$ is the minimal font-size (default 12px).
A seemingly reasonable alternative is the sum instead of the average. 
However, we found that the machine learned to delete texts to reduce costs which we did not want to allow for this issue.

\bpstart{Overlapping text}
The cost is computed as the sum of the total overlapping area. 
\begin{rev}
Using the sum we allow removing texts to solve overlapping, which is a common technique for responsive visualization design \cite{hoffswelltechniques}.
\end{rev}

\subsection{Heuristic}
\label{mvf:framework:algo}
We propose a greedy heuristic to train the agent effectively (Algorithm~\ref{algo:greedyheuristic}).
The heuristic is based on a policy-based approach that directly learns the optimal policy $\Pi(a | s)$.
The advantages of using a policy-based approach is the effectiveness in high-dimensional action spaces and the possibility to learn a stochastic policy  \cite{sutton2018reinforcement}, which adheres to the intuition that the ``optimal'' visualization is non-deterministic.

\setlength{\textfloatsep}{5pt}
\begin{algorithm}[t]
\SetAlgoLined
\SetKwInOut{Input}{Input}
\Input{Policy approximation $\Pi_{\theta}$, learning rate $\alpha$, penalty rate $\beta$}
\KwResult{hidden variable $\theta$}
 initialize $\theta$ with zeros\;
 Greedily select an existing mobile-friendly issue as current state $s_0$ and compute initial cost $J_0$\;
 i $\gets$ 0; $s \gets s_0$; $J \gets J_0$\;

 \While{$s$}{
 i++\;
 Sample an action $a_i$ based by $\Pi_{\theta}(a | s)$\;
 Evaluate the resulting state $s_i$, cost $J_i$, and return $R$\;
 $\theta \gets \theta + \alpha R \nabla_{\theta} log \Pi_{\theta} (a | s)$\;
  \If(\tcp*[h]{Current issue solved}){$s_i$ != $s$}{
   \If(\tcp*[h]{Deadlock}){$s_i$ is previously entered}{
       \ForEach{action $a$ between $s$ and $s_i$ }{%
 $\theta \gets \theta - \beta \nabla_{\theta} log \Pi_{\theta} (a | s)$\;

    }
   }
       $s \gets s_i$; $J \gets J_i$\;
   }
 }
 \caption{Greedy Heuristic}
 \label{algo:greedyheuristic}
\end{algorithm}

Specifically, \mvf is based on the well-established policy-gradient algorithm by Williams \cite{williams1992simple}.
This algorithm approximates the optimal policy using a parameterized function $\Pi_{\theta}(a | s)$, where $\theta$ is the hidden variable and $\Pi_{\theta}$ is usually the softmax function.
At each step, it updates the policy gradient through 
\begin{equation}
\label{policyGradient}
\theta \gets \theta + \alpha R \nabla_{\theta} log \Pi_{\theta} (a | s)
\end{equation}
where $\alpha$ is the learning rate, and $R$ is the return.

\begin{figure}[]
    \setlength{\abovecaptionskip}{0pt}
    \setlength{\belowcaptionskip}{5pt}
	\centering
	\includegraphics[width=1\columnwidth]{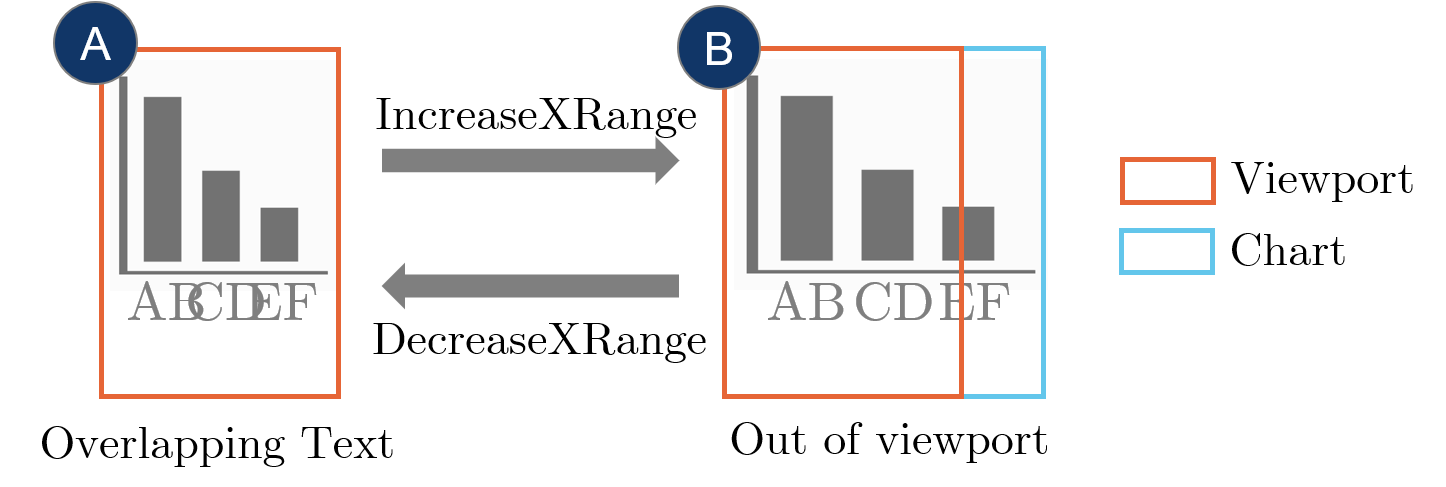}
	\caption{The naive greedy algorithm could take short sighted, selfish actions that cause deadlocks between two states A and B.}
	\label{fig:deadloop}
\end{figure}

\begin{figure*}[!t]
    \setlength{\abovecaptionskip}{-2pt}
    \setlength{\belowcaptionskip}{-5pt}
	\centering
	\includegraphics[width=1.9\columnwidth]{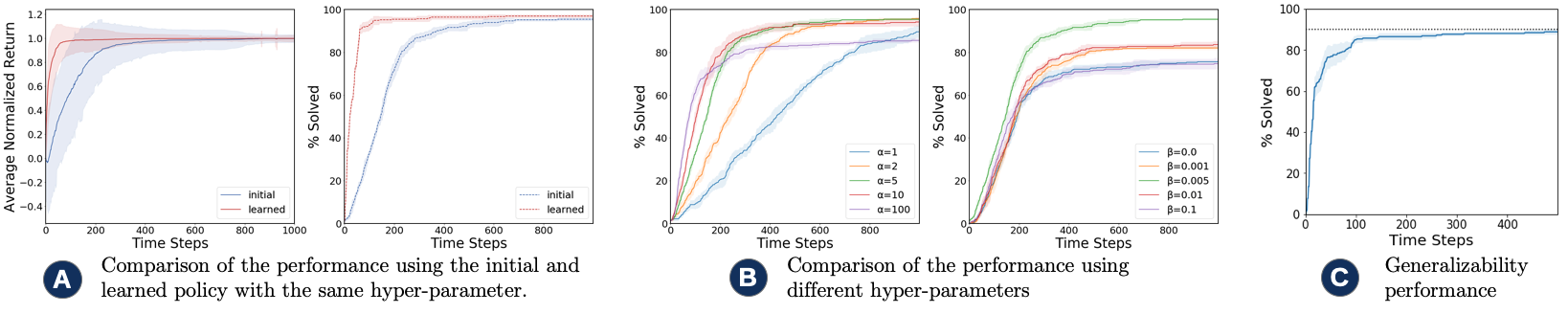}
	\caption{The solid line and shaded regions represent the mean and standard deviation respectively across 5 runs.}
	\label{fig:trained}
\end{figure*}

Here we simply use the reward $r$ at each timestamp as the return, that is, $R_{t} = r_{t}$, which allows us to update the policy at each timestamp.
The motivation is that one action could potentially have a significant influence on the visualization.
The reward $r_{t}$ is determined by the difference between the current and previous cost, normalized by the initial cost upon entering the state, since the cost function for each mobile-friendly issue varies from scales:
\begin{equation}
\label{reward}
r_{t} = \frac{{J_{t} - J_{t-1}}}{J_0}  
\end{equation}

\mvf\ utilizes a greedy algorithm to approximate the optimal solution.
\begin{rev}
It greedily selects the state based on a predefined order, as shown in the leftmost columns of Fig. \ref{fig:policy} (from top to bottom).
We experimentally define this order: global parameters (\textit{e.g.} margins, axes) are adjusted first, while font-related issues (\textit{e.g.}, font-size, text overlapping) are solved last.
\end{rev}
Thanks to the greedy nature, \mvf\ computes the cost and updates the hidden variable $\theta$ in respect to only the current state at each timestamp, thereby reducing the time complexity from $\mathcal{O}(|S|)$ to $\mathcal{O}(1)$ (\textbf{R5}).
However, such a greedy algorithm is ``short sighted'' -- it only focuses on solving the current mobile-friendly issue, potentially causing other issues.
\autoref{fig:deadloop} demonstrates an example where a deadlock occurs.

\mvf{} addresses this challenge by imposing a long-term penalty for deadlocks, which takes advantages of the fact that reinforcement learning is particularly well-suited to problems that include a long-term versus short-term reward trade-off \cite{sutton2018reinforcement}.
Upon entering a state $s_{i+1}$ that is previously visited, \mvf{} penalizes all actions from $s_{i}$ to $s_{i+1}$ by
\begin{equation}
\label{penalty}
\theta \gets \theta - \beta \nabla_{\theta} log \Pi_{\theta} (a | s)
\end{equation}
where $\beta$ is the penalty rate. 
In other words, \mvf{} imposes a long-term penalty to counterbalance the actions toward short-term payoffs.
As suggested by Mnih \textit{el al.}~\cite{mnih2013playing}, we use a fixed value which makes it easier to use the same rate across multiple visualizations.

\section{Evaluation}
\label{eva}
This section presents a series of quantitative and qualitative studies that aim to evaluate the performance, generalizability, and quality of the learned policy.
The core of \mvf{} is implemented in Typescript. 
We tested \mvf{} on a MacBook Pro 2015 with a 2.7GHz Intel Core i5 processor and 8GB memory.

\subsection{Training}
\begin{rev}
We trained the agent on a small dataset including $81$ visualizations.
The dataset was manually selected from our corpus to alleviate bias caused by an unbalanced data problem.
We kept charts that exhibits mobile-friendly issues and satisfies our prerequisites and removed charts with similar types, designs, and mobile-friendly issues.
\end{rev}

In supervised learning, it is usually straightforward to track the model's performance by evaluating the performance on the training and testing dataset.
However, accurately evaluating the progress of an agent in reinforcement learning is challenging \cite{mnih2013playing}.
Thus, we analyze the training performance using two metrics.
The first is the average cumulative reward (returns), which is the most commonly used of such metrics \cite{henderson2018deep}.
However, since the returns vary across different visualizations, we normalize the returns so that 100\% corresponds to the final score.
The second metric is the percentage of solved problems \cite{cobbe2018quantifying}.
Here we define that a visualization is solved if its cost is zero, \textit{i.e.}, no mobile-friendly issue is detected.
\begin{rev}
It should be noted that our cost functions exclude the distorted ratio issue,
which usually results into mobile versions with different aspect ratios from the desktop version.
\end{rev}

We perform 5 experiment runs with 1,000 time steps using the same hyper-parameter ($\alpha = 5, \beta = 0.005$). 
As suggested by Riedmiller \textit{el al.}\ \cite{riedmiller2007evaluation}, we compare the training performance of the initial policy and the learned policy.
As shown in \autoref{fig:trained} (A), the learned policy achieves rewards faster and with less variance.
Also, the learned policy speeds up problem solving and eventually has a slightly better solving rate.
Finally, \mvf{} successfully solves around 90\% of visualizations within 100 steps, and 95\% of visualizations within 1,000 steps.

We also investigate the impact of hyper-parameters, \textit{i.e.}, the learning rate $\alpha$ and penalty rate $\beta$, as demonstrated in \autoref{fig:trained} (B).
The results of the learning rate are consistent with existing knowledge \cite{grondman2012survey} -- small values (\textit{e.g.}, 1 and 2) require more training steps, while larger values (\textit{e.g.}, 100) cause the model to converge too quickly to a sub-optimal solution.
It should be noted that our values are considerably larger than typical values $(0, 1]$ to compensate for the percentage scaling caused by the normalization of the return in~\autoref{policyGradient}.

We observe high parameter sensitivity for the penalty rate, which controls the long-term penalty to counterbalance the greedy actions towards short-term rewards.
It tends to converge towards sub-optimal solutions when without penalty (\textit{i.e.}, $\beta=0$), which demonstrates the effectiveness of our penalty strategy.
\mvf{} turns out to be very robust when choosing a medium value $\beta=0.005$, while larger values lead to less successful learning behavior.
We also find that the convergence rate is not sensitive to the penalty rate. 
This might be because problem solving is mainly related to choosing ``good'' actions (\textit{i.e.}, get rewards) instead of avoiding ``bad'' ones (\textit{i.e.}, get penalized).
Specifically, penalizing ``bad'' actions does not necessarily contribute to increasing chances of ``good'' ones. 


\subsection{Evaluating Generalizability}
We further evaluate the performance of \mvf{} on another dataset to understand its generalizability across different visualizations. 
Since there is no benchmark dataset, we collected our test dataset by searching and crawling visualizations with the keyword ``Covid-19'' on Observable (https://observablehq.com) published in March 2020, with the intention of capturing recent practice in web visualization.
This results in 51 visualizations under the current prerequisite of \mvf{}.
We adopt the desktop-version chart as the input to demonstrate \mvf's generalizability across input specifications.

We follow the methods by Cobbe \textit{et al.}\ \cite{cobbe2018quantifying} to quantify generalization. 
Based on the results in the training, we perform 5 runs with 500 time steps on the testing dataset.
Fig.~\ref{fig:trained} (C) shows the performance.
In general, \mvf{} succeeds in finding good solutions in 89\% cases, which is slightly worse than that in the training dataset.

We identified two common causes for failing cases. 
First, the optimization problem can be ill-posed, \textit{i.e.}, there does not exist a solution given the actions defined by \mvf.
For instance, considering a vertical bar chart with more than 30 bars, possible solutions include removing some bars or changing the orientation to horizontal.
Future work should address how machines can reasonably take such actions.
Second, the training is insufficient since several states are seldom visited. 
In particular, \mvf{} performed poorly when marks were out of the viewport, which was not observed during training.
More training could alleviate this issue.



\subsection{Interpreting the Model}
\label{eva:interpret}
\begin{figure}[!t]
    \setlength{\abovecaptionskip}{0pt}
	\centering
	\includegraphics[width=0.9\columnwidth]{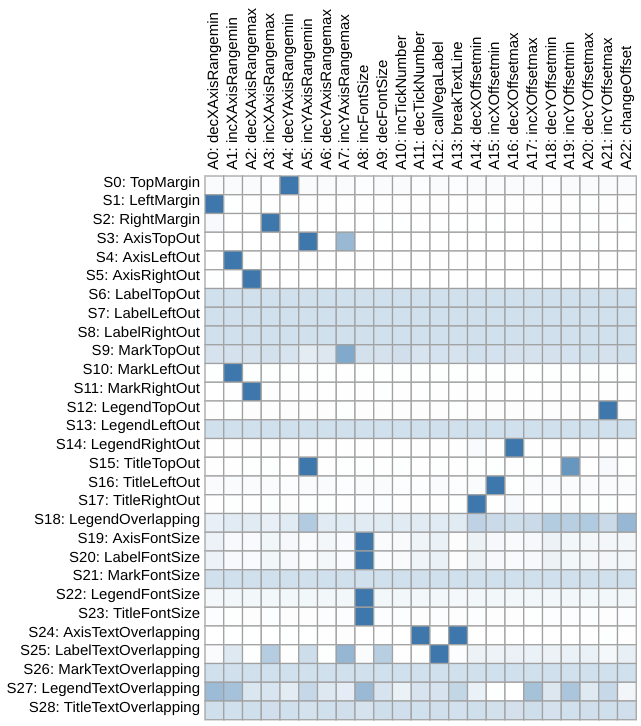}
	\caption{The learned policy - the probability that the agent takes an action (column) when in a state (row). The color encodes the probability (truncated to 0.3).}
	\label{fig:policy}
\end{figure}

We interpret the learned model to qualitatively evaluate the performance.
\begin{rev}
The learned policies can be interpreted as probabilistic decision rules.
As visualized in \autoref{fig:policy}, each row (S0-S28) corresponds to a state, and each column (A0-A22) represents an action.
The cell encodes the probability that the agent takes a decision (\textit{i.e.}, action) under a condition (\textit{i.e.}, state).
For instance, the first row indicates that when the top margin exceeds the threshold (S0), the agent has a near 100\% probability to decrease the min range of the y-axis (A4),
which conforms to human choices.
\end{rev}
In general, we identified four patterns, which provide implications for future improvements of automated approaches, as well as challenging issues to which human designers should attach importance when designing visualizations for mobiles.

\begin{rev}
First, the agent converges to a single action under most states.
For instance, when the margin exceeds the threshold (S0-2), the agent has learned to adjust the scale in corresponding orientations, which matches our intuition.
This shows that the agent has found a confident successful solution to this mobile-friendly issue.
\end{rev}
Second, the agent could also converge to multiple actions.
S24 illustrates such an example, where the agent has learned that both reducing the tick number (A11) and breaking lines (A13) can alleviate the text overlapping problem at axes. 
However, the latter has less chance of improvement.
Third, the agent does not get well trained on several states (\textit{e.g.} due to insufficient observations, S6-9, S21, S26), as it exhibits a near equal distribution of probabilities.
The former is due to the fact that labels were usually included by axes, that is, labels will stay in the viewport if the axes do.
The latter two seemingly indicate invalid states -- marks are unlikely to include text.
This suggests that states in \mvf{} can be further reduced.
\begin{rev}
Finally, the agent cannot make an action with high probability under fewer states (\textit{e.g.}, S18, S27), that is, the machine has difficulties solving those difficult issues.
This implies that designers should attach importance to certain challenging issues when creating web visualizations. 
\end{rev}

\subsection{Gallery}
\label{eva:case}
\autoref{fig:gallery} shows successful outputs of \mvf: given only the SVG as input, \mvf{} successfully generated  visualizations with improved mobile-friendliness while remaining faithful to the original information and style.
Note that there are subtle changes, e.g., the text labels in \autoref{fig:gallery} (D) are re-positioned. 
More examples can be found in the supplemental material along with illustrating videos.

\section{Discussion and Future Work}
We now reflect on \mvf{} and discuss areas for future work. 

\subsection{Interpretable Automated Visualization Design}
\begin{rev}
\mvf{} embraces algorithmic explainability and transparency that makes it easier for both model developers and end users to understand how the automatic system works.
\mvf{} supports play-backing the optimization process step-by-step to help users understand the automation process.
As described in \autoref{mvf:framework:model:cost}, explainability helps us reason about unexpected system behavior and distinguish seemingly reasonable cost functions during development.
More importantly, explainability allows interpreting our reinforcement learning model in the format of human-readable decision rules, which helps evaluate the quality of trained models.
Compared with existing rule-based approaches that are often deterministic, the learned rules are stochastic and thus able to generate more flexible solutions to diverse visualizations, while removing the heavy manual effort of writing and polishing rules that scale well to real-world diversity.

With this in mind, \mvf{} adds to the recent discussion on rule-based and model-based systems for automatic visualization design \cite{saket2018beyond}.
While model-based methods (\textit{e.g.}, \cite{dibia2019data2vis, hu2019vizml}) have demonstrated promising results, 
they have not proven definitely superior to carefully crafted rules derived from human knowledge.
However, rule-based approaches face limitations such as expensive rule creation and the combinatorial explosion of possible conditions \cite{aggarwal2016recommender}.
As such, \mvf{} demonstrates a hybrid perspective that augment rules with models for their combined power.
\mvf{} automatically learn reasonable decision rules from a reinforcement learning model, given that human-crafted rules in existing systems have not scaled to diverse real-world conditions.
In the future, we are excited to explore how to embed rules in models to leverage the advantages that rules allow flexible and continuous extension.
One possible solution is to enable users to adjust the policy and add customized states or actions.
\end{rev}

\subsection{Improving \mvf{}}
\begin{rev}
\mvf{} has several limitations and further work is warranted to improve its usability.

\textbf{Overcoming simplifying assumptions.}
\mvf{} does not address interaction problems since most web visualizations are static~\cite{hoffswelltechniques}.
Future work should study how to automatically model and deconstruct interactive visualizations.
Besides, \mvf{} omits the perception problems, 
since existing perceptual studies (e.g.,\ \cite{brehmer2019comparative,brehmer2018visualizing}) on mobile visualizations only focus on a limited set of chart types.
Future research should address a wider range of visualization designs.
\end{rev}

\textbf{Balancing agency and automation.}
We are excited to extend \mvf{} to include humans in the loop. 
For example, \mvf{} does not currently enable designers to input dependency relationships between non-layout properties among visual elements, such as the sizes of titles and axis labels, or to choose a desired aspect ratio to avoid distortions.
It would be interesting and valuable to design and develop an interactive, semi-automated tool by taking a human-in-the-loop approach (\textit{e.g.},~\cite{jung2017chartsense}),
\textit{i.e.}, supporting manual adjustment based on automatically generated results. 
\textbf{Improving greedy heuristics.}
\mvf{} utilizes problem reduction techniques and a greedy heuristic to solve a complex, multi-criteria optimization problem which may not find a ``best'' solution.
For instance, our parameter specification might not accurately represent some visualizations, especially those with deliberate data- or context-dependent human design choices.
Like similar greedy algorithms, \mvf{} speeds up computation at the cost of not always converging to a global optimum.
Future research should propose approaches that better address this trade-off.
\begin{rev}
Moreover, the greedy heuristic utilizes a pre-defined order to solve the multi-objective optimization problem, which could result into sub-optimal results.
We hope to improve the performance by applying more advanced techniques such as adaptive reinforcement learning to dynamically update the greedy state.
\end{rev}


\textbf{Quantifying the mobile-friendliness through empirical studies.}
We evaluate \mvf{} through quantitative studies on two datasets containing 132 real-world visualizations.
However, due to the scarcity of benchmarks, we cannot conclusively determine that our datasets are fully representative.
\begin{rev}
Besides, we do not consider the distort ratio issue since resizing visualizations is a common technique, 
which, however, might cause perceptional bias that warrants future empirical studies.
\end{rev}
Our approach also inherits limitations common to reinforcement learning that the evaluation metric is based on the training objectives.
Our results show that \mvf{} could efficiently solve 89\% cases of the optimization problem with respect to the defined cost functions. 
However, it does not directly reflect on the overall quality of generated results because there lacks a metric for measuring mobile-friendliness for visualizations, which is currently not supported by Google and Bing mobile-friendly test tools.
\begin{rev}
In the future, we plan to study such metrics through user studies and evaluate \mvf{} with those metrics on more data.
\end{rev}

\acknowledgments{
The authors wish to thank A, B, and C. This work was supported in part by
a grant from XYZ (\# 12345-67890).}

\bibliographystyle{abbrv-doi.bst}

\bibliography{main}
\end{document}